\documentclass{aims}
\usepackage{amsmath}
  \usepackage{paralist}
  \usepackage{graphics} 
  \usepackage{epsfig} 
 \usepackage[colorlinks=true]{hyperref}
\hypersetup{urlcolor=blue, citecolor=red}

\def\currentvolume{X}
 \def\currentissue{X}
  \def\currentyear{200X}
   \def\currentmonth{XX}
    \def\ppages{X--XX}

\usepackage{amsrefs,ulem}
\newcommand{\RR}{\mathbb{R}}
\newcommand{\ra}{\rightarrow}
\newcommand{\ud}{\mathrm{d}}

\title[Predicting Drug Release Kinetics]{Predicting the Drug Release Kinetics of Matrix Tablets}
\author[B. Baeumer, L. Chatterjee,  P. Hinow, T.  Rades, A. Radunskaya and I. Tucker]{}
\author[B. Baeumer \textit{et al.}]{}

\email{hinow@ima.umn.edu}
\email{aradunskaya@pomona.edu}

\thanks{The workshop that initiated this work was supported by NSF grant DMS-0737537.  PH is supported by an IMA post-doctoral fellowship.}

\begin{document}
\maketitle

\centerline{\scshape Boris Baeumer }
{\footnotesize
 \centerline{Department of Mathematics and Statistics, University of Otago }
   \centerline{Dunedin, New Zealand}
} 

\medskip

\centerline{\scshape Lipika Chatterjee }
{\footnotesize
 \centerline{ New Zealand's National School of Pharmacy}
   \centerline{University of Otago, Dunedin, New Zealand}
} %

\medskip

\centerline{\scshape Peter Hinow}
{\footnotesize
 \centerline{ Institute for Mathematics and its Applications, University of Minnesota}
   \centerline{ 114 Lind Hall, Minneapolis, MN 55455, USA}
} %
\medskip

\centerline{\scshape Thomas Rades }
{\footnotesize
 \centerline{ New Zealand's National School of Pharmacy}
   \centerline{University of Otago, Dunedin, New Zealand}
} %
\medskip

\centerline{\scshape Ami Radunskaya}
{\footnotesize
 \centerline{ Department of Mathematics, Pomona College}
   \centerline{610 N. College Ave., Claremont, CA  91711, USA}
} %
\medskip

\centerline{\scshape Ian Tucker}
{\footnotesize
 \centerline{ New Zealand's National School of Pharmacy}
   \centerline{University of Otago, Dunedin, New Zealand}
} %

\bigskip

\begin{abstract} In this paper we develop two mathematical models to predict the release kinetics of a water soluble drug from a polymer/excipient matrix tablet. The first of our models consists of a random walk on a weighted graph, where the vertices of the graph represent particles of drug, excipient and polymer, respectively. The graph itself is the contact graph of a multidisperse random sphere packing. 
The second model describes the dissolution and the subsequent diffusion of the active drug out of a porous matrix using a system of partial differential equations. The predictions of both models show good qualitative agreement with experimental release curves. The models will provide tools for designing better controlled release devices.

\end{abstract}
\section{Introduction}\label{section:introduction}
\begin{sloppypar}
The first Workshop on the Application of Mathematics to Problems in Biomedicine took place from December 17-19, 2007 at the University of Otago in Dunedin, New Zealand. During that workshop, our group of nine, comprising pharmaceutical scientists and  mathematicians worked on the problem of predicting the drug release kinetics from a sustained release matrix tablet. The present paper grew out of this event. 
\end{sloppypar}

Sustained release (also called extended release) tablets are a common dosage form. A sustained release (SR) tablet is typically designed to release drug over 12-24 hours and might contain three times the dose of drug that is contained in an immediate release tablet. In this way, a patient need take a tablet only once a day rather than three times a day if immediate release tablets were used.  This not only has the advantage of convenience for the patient but ideally provides more constant levels of drug in the body. Fluctuating drug levels can result in the patient being exposed to  levels of drug which are too high at  times, leading to harmful side-effects,  and sub-therapeutic levels at other times. Sustained release tablets can smooth these fluctuations leading to better control of the patient's illness or symptoms.

Drugs formulated in sustained release tablets typically have half-lives in the body of 3-8 hours. A drug with a long half-life given as a standard tablet once or twice a day inherently maintains reasonably constant levels in the body making sustained release tablets unnecessary. A drug with a very short half-life (say $1\,h$) would have to be given many times each day if formulated as a standard tablet. This might then suggest that an sustained release tablet would be useful, but such a tablet would have to contain perhaps 12 standard doses, and this would be potentially dangerous. 

Sustained release tablets can be formulated in various ways but the preferred approach, because of its apparent simplicity and ease of manufacture, is to make a sustained release matrix tablet. A matrix tablet can be made by mixing the drug with suitable excipients (non-active components of the formulation) and compressing the mix in a die at high pressure, say $100\,MPa$,  thereby producing a tablet. Excipients are added to tablets for various reasons. Some are diluents to increase bulk and aid compaction. Others help in manufacture, for example lubricants and flow enhancers, while others influence the behavior of the tablet in water. For example, disintegrants added to standard tablets making them break-up when placed in water, are not used in sustained release tablets designed to remain intact. Excipients vary in their physical properties, such as water solubility and mechanical properties.

An important excipient in sustained release matrix tablets is the agent to control the drug release. Broadly, either lipids, such as fats and waxes, or polymers are used. In this paper we will further divide excipients into the insoluble polymer which forms the coherent matrix and soluble excipients. Some polymers, such as hydrogels, swell on exposure to water and create a rate-controlling gelatinous layer around the SR tablet. Other polymers, as used in the tablets described below, do not swell markedly but create a matrix through which water can penetrate to dissolve the drug particles and any water-soluble excipient particles. The dissolved particles then exist as molecules which diffuse down the concentration gradient, through the water-filled pores, to be released from the tablet. Several of the authors have been studying matrix tablets made by mixing drug with a polymer and one additional excipient, either lactose, a water-soluble brittle excipient whose particles might fracture during compression, or mannitol, a water-soluble plastic excipient whose particles plastically deform during compression. Other factors can potentially affect certain characteristics such as  the hardness or sustained release behavior of the tablet. These include the particle size of the excipients, proportions of ingredients, compression force and post-compression thermal treatment. The latter factor can change the table porosity due to coalescence of polymer particles to form a coherent polymer network \cites{Krajacic, Azarmi, Chatterjee}. The amount of drug in the tablet is determined by the total  dose to be given to the patient, however, the proportions of the excipients vary with their function and the specific tablet formulation. For example, a lubricant such as magnesium stearate might be used at a 1\% level, whereas the level of a rate-controlling polymer would be much higher (10-50\%) but preferably the level would be kept as low as possible for cost reasons. 

\begin{figure}[th]
\begin{center}
\includegraphics[width=55mm, height=48mm]{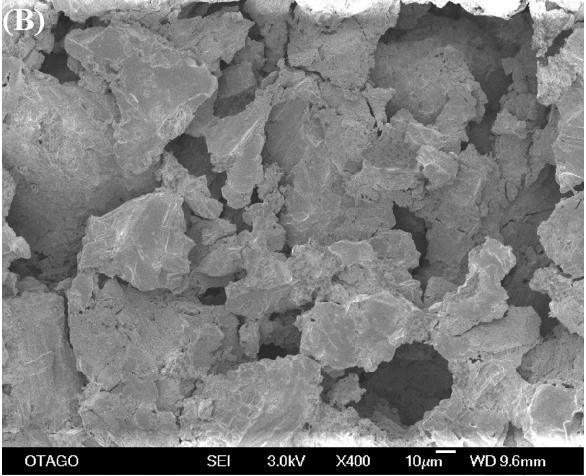}
\includegraphics[width=55mm, height=48mm]{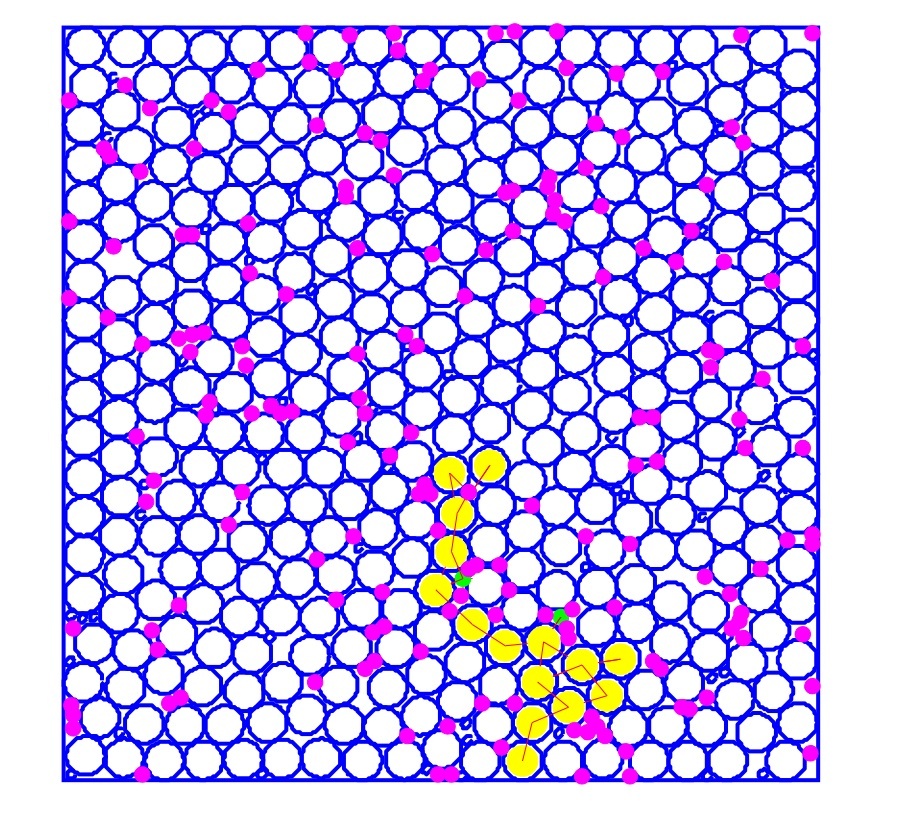}
\caption{\textit{Left panel:} Field emission scanning electron microscope (SEM) image of a matrix tablet after dissolution of drug and excipient at $400\times$ magnification. \textit{Right panel:} A sample path through a schematic two-dimensional tablet from an inner particle to the edge.  Small solid circles represent polymer particles.  Notice that there are multiple paths from each initial particle location.}
\label{Fig:WalkOnGraph}
\end{center}
\end{figure}

When such tablets are placed in water, or in the intestinal fluid of the patient, the coherent polymer matrix sustains release of the drug by keeping the tablet largely intact and by providing a tortuous network through which water penetrates and dissolved drug and soluble excipient molecules diffuse out. That is, the postulated release mechanism involves penetration of fluid, dissolution of the drug and soluble excipient in the fluid, and outward diffusion of molecules of dissolved drug  due to the concentration difference between the solution in the tablet and the intestinal fluid. Once released from the tablet, the drug is rapidly absorbed through the patient's intestine into the blood stream. Ideally, the delivery rate of the drug from the tablet should be such as to maintain the preferred blood level for an extended time. The left panel of figure \ref{Fig:WalkOnGraph} shows the structure of the polymer matrix of a tablet after dissolution of the solvable excipient and the drug.

In this paper we propose two mathematical models that  predict the release kinetics of matrix tablets, taking into account parameters such as the composition of the powder mixture, powder particle sizes and applied curing temperature.  The first model models the diffusion process as a random walk through the channels in the tablet that are formed in the compaction and heating process.  In this  model, the particle distribution within the tablet is represented by a graph  embedded in ${\mathbb R}^3$ whose vertices are generated from a random sphere packing and in which two vertices are joined by an edge if the corresponding spheres are in close proximity. On this contact graph  we perform random walks that start at the location of each drug particle. These random walks mimic the diffusion of the drug molecules from their initial position to the edge of the tablet.  By adjusting the probabilities of moving to vertices representing different types of particles, the composition of the tablet affects the length distribution of these exit paths in a realistic way.
The right panel of figure \ref{Fig:WalkOnGraph} shows a sample path through a random packing of disks.

Low levels of drug in some polymer systems can lead to trapping and incomplete release. Percolation theory has been applied to understand these processes \cites{Bonny91,Bonny93,Caraballo93,Caraballo96,Leuenberger,ElArini}.  Monte Carlo simulations of drug release from a binary powder mixtures with inert (water insoluble) excipient have been performed by Villalobos  \textit{et al.}~\cites{Villalobos05,Villalobos06}. The drug and excipient particles were placed on the sites of a cubic lattice and the drug particles performed random walks on the accessible empty sites under a volume exclusion constraint. The authors reported different exponents for the release profiles as function of time, with $\sqrt{t}$-like behavior for large drug fractions. In our work we take into account drug, excipient and polymer, where the polymer plays the role of the inert matrix. More importantly, we also allow for differences in particle sizes and irregular placements of the particles in space. 

The second model investigates a different approach to predicting release kinetics of tablets using partial differential equations,  where time and space are treated as continuous variables. Previous work using this approach includes \cites{Lemaire, Siepmann}. Suitable versions of the diffusion equation (often in cylindrical coordinates, resembling the natural shape of a tablet) are formulated for the diffusion of dissolved drug and (e.g. in \cite{Siepmann}) water penetrating the tablet. These models can take into account the dissolution kinetics of the drug in different types of buffer. Siepmann and Peppas \cite{Siepmann} consider hydrophilic polymer matrices that swell and then dissolve, unlike the polymers used in the tablets that we are studying.  While these results may not be directly comparable to our results for inert matrices, they do show that an increasing drug load results in faster release kinetics \cite{Siepmann}*{Fig.~4}. 

\section{Model 1: The diffusion of drug molecules is a random walk on a weighted graph}\label{section:model_1}

This first model is constructed in three distinct steps. In the first step, the positions of the particles in the tablet are taken from a random sphere packing, where each particle is a sphere of a fixed radius. In Step 2, this random sphere packing is represented as a graph embedded in ${\mathbb R}^3$, whose vertices are the particles, with edges between particles that are close to each other in space. The compression of the tablet is modeled as a deformation of this graph, and the heating of the tablet is modeled by removing edges. Finally, in Step 3, the diffusion of the drug particles to the exterior of the tablet is modeled as a random walk on the graph generated in Step 2. Drug release profiles are then generated as the distribution of exit times of these particles, using the following argument. 

Since we model the diffusion process as a random walk, the distance that a particular drug molecule must traverse from its position inside the tablet to the exterior is a random variable, $L$.  Let $f_L(s)$ denote the density of these path lengths. Assume that a dissolved drug molecule performs Brownian motion with diffusion constant $D$ along these paths.   If the particle has to travel along a path of length $s$ to get to the exterior of the tablet,  then the travel time $T$ is a random variable with  density function (see \cite{Feller}*{VI.2})
\begin{equation*}
f_{T|s} (\tau) = \frac{s}{\sqrt{2 \pi D \tau^3}} e^{\frac{-s^2}{2D\tau}}. 
\end{equation*}
In other words, $f_{T|s}$ is  the density of $T$ conditioned on the path-length being equal to $s$.  Thus, we can integrate this conditional density against the density of path lengths to get the density of escape times of the particles from the tablet
\begin{equation*}
f_T(\tau) =  \int_0^\infty f_{T|s}(\tau) f_L(s) \,\ud s = \int_0^\infty \frac{s}{\sqrt{2 \pi D \tau^3}}
e^{-\frac{s^2}{2D\tau}} f_L(s) \, \ud s.
\end{equation*}
It's cumulative distribution function  
\begin{equation*}
 F_T(t) = \int_0^t f_T(\tau) \, \ud \tau
\end{equation*}
gives the release profile of the matrix tablet. In Step 3, we estimate $F_T(t)$ by performing many random walks on the graph created in Step 2.

We now describe each of these three steps in more detail.  Results from preliminary implementations of this model are given in Section \ref{section:results}.  In this work we are not interested in supplying the optimal algorithm for completing each step in the modeling process.  Rather, we are trying to validate the idea and, if the results appear sound, we can then optimize the algorithms.  We assume that drug, excipient and polymer particles can be modeled as perfect hard (i.e.~non-deformable) spheres, where all particles of each type have the same radius, but the three radii of the three different types of particles may differ from one another. The assumption that all particles of one species have the same radius is a simplification. In practice,  their sizes are distributed around a mean with a given variance. Relaxing this assumption, as well as including particles of other, non-spherical shapes are topics of future research. 
 
{\bf Step 1:}  We assume that prior to compaction the powder mixture is a dense multidisperse random sphere packing. Dense packings of disks or spheres have attracted the interest of mathematicians, physicists and engineers for a long time, see e.g.~\cites{Torquato,Talbot,Lubachevsky,Donev,Knott} and the references therein. It turns out to be extremely difficult to give a precise mathematical meaning to the concept of a ``dense random sphere packing'', as the desired properties of high density and high degree of randomness conflict with each other  \cite{Torquato}. The more dense a packing is, the more ordered it tends to be, culminating in the highly ordered lattice packings such as the tridiagonal packing of disks in $\RR^2$ and the face-centered cubic packing of balls in $\RR^3$. (The optimality of the latter packing was conjectured by Johannes Kepler in 1611 and proved by Thomas Hales in 1998.) While this question is certainly interesting from a mathematical point of view, we shall not pursue it further. Rather, we will work with the outcome of an experimental protocol that produces jammed configurations of hard spheres. 
 
The protocol used in this paper was first suggested by Lubachevsky and Stillinger \cite{Lubachevsky} and later expanded by Knott and coworkers \cite{Knott}. The paper by Knott \textit{et al.}~\cite{Knott} discusses the problem of packing spheres of oxidizer and fuel in a solid rocket propellant. Briefly, the Lubachevsky-Stillinger (LS) protocol proceeds as follows. A random initial configuration of sphere centers $(\boldsymbol{x}_1(0),\,\boldsymbol{x}_2(0),\,\dots\,\boldsymbol{x}_N(0))$ and a random set of initial velocities $(\boldsymbol{v}_1(0),\,\boldsymbol{v}_2(0),\,\dots\,\boldsymbol{v}_N(0))$ are chosen. The spheres are partitioned into $M$ radius classes, each class contains $N_i$ spheres ($i=1,\dots,M$) and $N=\sum_{i=1}^MN_i$. The spheres move within a container and collide with each other and with the walls of the container. As time evolves, the radius of class $i$ spheres grows according to $r_i(t)=a_it$. Thus the ratios $r_i(t)/r_j(t)$ are preserved throughout the process. An event is a collision of two spheres or the collision of a sphere with the wall. At each collision of two spheres the velocities of the colliding spheres are updated in such a way that the spheres move away from each other after the collision.  This requires an increase in mechanical energy since the radii are growing at a constant rate. Eventually, as the packing becomes more and more dense, the time between two collisions approaches zero and the process is stopped. The resulting configuration of spheres is (nearly) rigid, in the sense that no sphere can be moved while keeping the centers of all other spheres fixed. For further details we refer to  \cites{Lubachevsky,Knott}. It should be noted that there exist alternative methods to construct random dense sphere packings \cites{Jodrey,Tobochnik,Stepanek}.

\begin{figure}[th]
\begin{center}
\includegraphics[width=55mm,height=48mm]{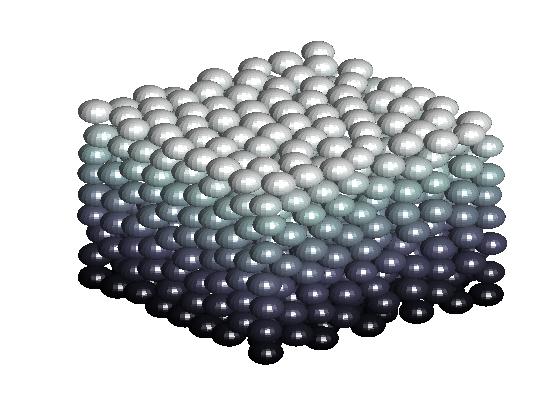}
\includegraphics[width=55mm,height=48mm]{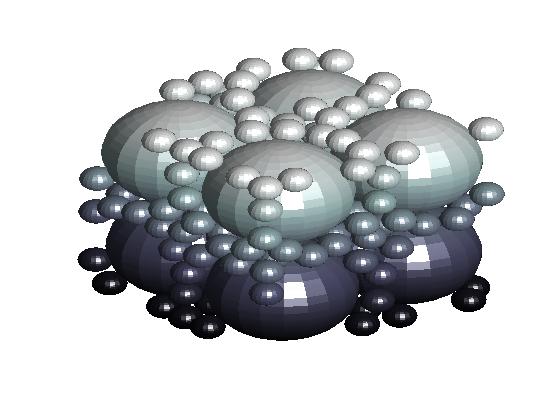}
\caption{\textit{Left panel:} The result of the Lubachevsky-Stillinger protocol using a total of 735 spheres where drug, polymer and excipient particles are present in a mixture $90:100:545$. The excipient particles are slightly larger than the others (see table \ref{Tab2}). The packing density is approximately $0.54$. \textit{Right panel:} The result of the Lubachevsky-Stillinger protocol using a total of 198 spheres where drug, polymer and excipient particles are present in a mixture $90:100:8$. The excipient particles are about 4 times bigger than the others (see table \ref{Tab2}). The packing density is approximately $0.63$. }\label{3D_pack}
\end{center}
\end{figure}

{\bf Step 2:}
Having obtained a multidisperse dense random packing of spheres, this sphere packing is compressed as the powder is compressed in the die \cite{Stepanek}. We model the die as a right circular cylinder whose axis is the $z$-axis and whose lower surface lies and remains in the $xy$-plane. The top surface of the cylinder which was originally in the plane $\{z=1\}$ is displaced by $\alpha\le 1$. A particle originally located at the level $z$ is displaced to the level $\alpha z$.  Using the new coordinates, a graph $G$ is constructed, the proximity graph of the compacted sphere packing. This is a Euclidean graph $G = \{V, E\}$ whose vertices are the compressed sphere centers. Each vertex carries a label $L:V\ra\{D, P, X\}$ that indicates whether the sphere is a drug, polymer or excipient particle, respectively. Two vertices are connected by an edge if the corresponding spheres are sufficiently close to each other. We also introduce a surface indicator  $s:V\ra\{0, 1\}$ where  $s(v)=1$ if the sphere centered at $v$ is in contact with the exterior of the tablet.

Next, we form the \textit{heated contact graph} $\hat{G}$. If the centers of two spheres are connected by an edge, this edge is removed with a probability $p$ that is an increasing function of the heating temperature $T_h$ and the duration of the heating process $t_h$. This reflects the fact that polymer particles will amalgamate and block the access of solvent to certain particles. In the extreme case we observe drug or excipient particles that are completely surrounded by a coat of molten polymer particles. This results in a certain amount of drug that is not released within a physiologically relevant time span. Such behavior has been observed in experimental release curves, see figure \ref{exp_data}.

{\bf Step 3:}
The estimation of the distribution of path lengths, $f_L(t)$,  and the modeling of the diffusion process on these graphs is streamlined in this implementation. Theoretically, each molecule undergoes one dimensional Brownian motion along a path from its initial location until it reaches the edge of the tablet.The one-dimensional diffusion can be written as a limit of one-dimensional
random walks, where the step size, $\Delta x$, and the time between steps,
$\Delta t$, go to zero in such a what that, in the limit, the variances do
not tend to either zero or infinity. This requirement is met by requiring
that
\begin{equation*}
\lim_{\Delta x, \Delta t \rightarrow 0} \frac{(\Delta x)^2}{\Delta t} = 2d,
\end{equation*}
where $d$ is the {\it diffusion constant}, see \cite{Feller}*{XIV.6, p.~325}. As a coarse approximation to the one-dimensional diffusion, therefore, we allow the molecules to perform random walks along the vertices of the heated contact graph $\hat{G}$, starting at every vertex of type $D$. These random walks approximate the diffusion process that each drug molecule undergoes in order to exit the tablet, and therefore give us an approximation of the distribution of exit times, $f_T$.  In practice, the physical length of each step in the random walk is related to the diameters of the spheres representing the original particles, hence the actual exit time is proportional to the sum of the lengths.  In the simulations used to generate the distribution of exit times shown in figures \ref{low_temp} and \ref{high_temp}, the radii of all the original particles were approximately the same, hence, in this case, the number of steps in the random walk was used as a surrogate for the time to exit the tablet.  Further work must be done to investigate the scaling properties of these random walks, as well as the effect of including particles of very different sizes.  However, our preliminary investigations indicate that the distribution of the number of steps in the walks is at least qualitatively the same as the distribution of exit times.  

The effect of the polymer, in particular the effect of the fusing of polymer particles in the heating process, is modeled by adjusting the probability of traversing a particular edge in the contact graph. Different rules can be implemented and tested. The simplest is to assign a conductivity $c_{ij}$ to the edge that connects vertices $v_i$ and $v_j$.  The conductivity is large for edges that connect vertices of type $D$ or $X$,  and small for an edge to or from a particle of type $P$.
Let  $\mathcal{N}(i)$ be the set of vertices neighboring vertex $i$, $v_i$.  Note that $ i \in \mathcal{N}(i)$,
so a molecule can remain at its initial position. If the walker, representing a drug molecule,  is situated at vertex $v_i$ then the probability of moving to the adjacent vertex $v_j$ is given by 
\begin{equation*}
p(i\ra j) = c_{ij}\left(\sum_{j\in\mathcal{N}(i)} c_{ij}\right)^{-1}.
\end{equation*}
The walk is stopped if either an exterior vertex or a prescribed maximum number of steps $N_{max}$ is reached. Notice that the metric on  $\hat{G}$ is the Euclidean distance between its vertices inherited from its embedding in ${\mathbb R}^3$, so that the distance a molecule travels on the graph corresponds to the actual distance it must travel in space. If a random walk does not reach one of the exterior vertices within $N_{max}$ steps, then the molecule is considered trapped inside the tablet.  This will be used to explain partial release of drug at higher polymer concentrations.  

\section{Model 2: The continuous model}\label{section:model_2}

In the second model we regard time and space as continuous
quantities and we set up a system of partial differential equations
for the contents of dissolved and undissolved excipient and drug in
the tablet. Our starting point is the assumption that fluid quickly
permeates the tablet \cite{Lemaire} and that dissolution is limited
by the saturation of the fluid with excipient and drug. This causes
a delay in the initial release of drug molecules until porosity
increases through diffusion at the boundary. This delay can be
observed as a change in concavity near $t = 0$ in the experimental
release profiles, see figure \ref{exp_data}.

Let $\Omega$ be the spatial domain of the tablet. We introduce cylindrical coordinates $(r,\theta,z)$ such that
\begin{equation*}
\Omega = \{ (r,\theta,z)\::\: 0\le r\le R, \, 0\le z\le H, \, 0 \le \theta < 2 \pi \}.
\end{equation*}
and assume that our tablet dissolves symmetrically,
i.e.~concentrations do not depend on the angular variable $\theta$
\cite{Siepmann}. Let $u_1$ be the concentration of dissolved
excipient in the solvent. Let $u_2$ be the concentration of
undissolved excipient in the solid remainder of the tablet.
Likewise, we denote by $v_1$ and $v_2$ the concentration of solved
drug in the solvent and the content of undissolved drug in the
tablet, respectively. All these quantities have dimension $ML^{-3}$.
By $\kappa$ we denote the porosity of the tablet, $\kappa$ is a
dimensionless value between $0$ and $1$ and will increase as more
and more excipient and drug are dissolved in the solvent. Then
$\kappa u_1$ is the concentration of solved excipient in the tablet.
We model the diffusion classically by assuming that the flux of
solved excipient across an interface is proportional to the
concentration gradient times the area of the interface. The area of
an interface is also proportional to $\kappa$, i.e.~the flux is
given by
\begin{equation}\label{Flux}
  Flux_{\mbox {solved excipient}}=-D_u\begin{pmatrix}
  \kappa \frac{\partial}{\partial r} u_1(r,z)\\
  \kappa\frac{\partial}{\partial z} u_1(r,z)\end{pmatrix},
\end{equation}
where $D_u$ is the diffusion constant of the dissolved excipient
(and  $D_v$ is the diffusion constant of the dissolved drug). The
conservation of mass equation then yields
\begin{equation*}
\begin{aligned}
\frac{\partial}{\partial t}(\kappa u_1) &= \frac
1r\frac{\partial}{\partial r}\left(rD_u \kappa
\frac{\partial}{\partial r} u_1(r,z)\right) +
\frac{\partial}{\partial z}
\left(D_u \kappa\frac{\partial}{\partial z} u_1(r,z)\right)+g(u_1,u_2)\\
&=: \nabla_{(r,z)}\cdot (D_u \kappa\nabla u_1) + g(u_1,u_2),
\end{aligned}
\end{equation*}
where $g(u_1,u_2)$ is the rate of concentration increase from dissolving
excipient. The higher the porosity, the more contact there is
between the solid excipient and the fluid in the pore space and
hence the rate of dissolution increases with porosity. We assume
this to be linear for simplicity but plan to explore more complex
relationships in future work. A similar modeling assumption has been
made by Lemaire \textit{et al.}~\cite{Lemaire}, see in particular
equation (1) in that paper. The higher the concentration of
dissolved excipient, the slower the rate of dissolution with no
dissolution taking place at the saturation point $C^u_{\max}$ of the
fluid. Hence we assume that
\begin{equation}\label{dissolution_kinetics}
    g(u_1,u_2)=\alpha_u \kappa \left({ 1 - \frac{u_1}{C^u_{\max}} }\right)u_2
\end{equation}
and we obtain the following  system of evolution equations
\begin{equation}\label{uv_system}
\begin{aligned}
&\frac{\partial}{\partial t}(\kappa(u_2,v_2) u_1) - \nabla_{(r,z)}\cdot(D_u \kappa(u_2,v_2)\nabla u_1) = \alpha_u \kappa(u_2,v_2) \left({ 1 - \frac{u_1}{C^u_{\max}} }\right)u_2, \\
&\frac{\partial}{\partial t} u_2 =-\alpha_u  \kappa(u_2,v_2) \left({ 1 - \frac{u_1}{C^u_{\max}} }\right)u_2 , \\
&\frac{\partial}{\partial t}(\kappa(u_2,v_2) v_1) - \nabla_{(r,z)}\cdot(D_v \kappa(u_2,v_2)\nabla v_1) = \alpha_v  \kappa(u_2,v_2) \left({ 1 - \frac{v_1}{C^v_{\max}} }\right)v_2 , \\
&\frac{\partial}{\partial t} v_2 =-\alpha_v  \kappa(u_2,v_2) \left({
1 - \frac{v_1}{C^v_{\max}} }\right)v_2.
\end{aligned}
\end{equation}
The rates of dissolution of excipient and drug are denoted by
$\alpha_u$ and $\alpha_v$, respectively. The porosity
$\kappa(u_2,v_2)$ depends on the concentration of undissolved
excipient and drug in the tablet. We assume that
\begin{equation*}
\kappa(u_2,v_2) =\kappa(u_2+v_2)=(\kappa_{0}-\kappa_{end})\frac{u_2+v_2}{u_{2}^0+v_{2}^0}+\kappa_{end},
\end{equation*}
where $\kappa_{0}$ is the initial porosity and $\kappa_{end}$ is the
porosity of the tablet once all the excipient and drug are
dissolved. The initial concentrations of undissolved excipient and
drug are denoted by $u_2^0$ and $v_2^0$. We assume that initially no
excipient and drug are dissolved and that the solid excipient and
drug are uniformly distributed,
\begin{equation*}
 u_1^0 = v_1^0 = 0, \quad u_2^0(x)\equiv u_2^0, \quad \textrm{and} \quad v_2^0(x)\equiv v_2^0,
\end{equation*}
for positive constants $u_2^0$ and $v_2^0$. If drug and excipient
are completely undissolved then the initial porosity $\kappa_0$ will
be about $2\,\%$, that is the porosity after compaction. If drug and
excipient are completely dissolved then the porosity will be about
$\kappa_{end}=60\,\%$. Equations \eqref{uv_system} are completed by
homogeneous Dirichlet boundary conditions for $u_1$ and $v_1$
\begin{equation*}
 u_1 = 0, \quad  v_1 = 0
\end{equation*}
on $\partial\Omega$, as we assume that any dissolved excipient or  drug outside the tablet is immediately carried away.

We add the first two equations of system \eqref{uv_system},
integrate over $\Omega$ and apply the divergence theorem. We obtain,
after exchanging the order of integration and differentiation
\begin{equation*}
\frac{d}{dt}\int_\Omega (\kappa(u_2+v_2) u_1+u_2)\,\ud x =
 D_u \int_{\partial \Omega}  \kappa(u_2+v_2) \nabla u_1 \cdot \mathbf{n}\,\ud \sigma =: -J_u(t),
\end{equation*}
where $\mathbf{n}$ denotes the outward normal and $\sigma$ the surface
measure.  This is the rate of change of the excipient load inside
the tablet and defines the flux of excipient $-J_u(t)$ across the
boundary $\partial \Omega$ of the tablet. (Alternatively we could
integrate the normal component of the flux in \eqref{Flux} over the
boundary). A similar calculation gives the flux of drug
\begin{equation*}
-J_v(t) = D_v \int_{\partial \Omega} \kappa(u_2+v_2) \nabla v_1
\cdot \mathbf{n}\,\ud \sigma.
\end{equation*}
The cumulative amount of drug released is then given by
\begin{equation*}
R_v(t) =\int_0^t J_v(s)\, \ud s.
\end{equation*}
The solutions $u_1,\,u_2,\,v_1$ and $v_2$ remain positive and
\begin{equation*}
\lim_{t\ra\infty} \big(||u_1(t)||_\infty+||u_2(t)||_\infty+||v_1(t)||_\infty+||v_2(t)\|_\infty\big) = 0.
\end{equation*}
It follows then that
\begin{equation}\label{pde_complete_release}
\lim_{t\ra\infty} R_v(t) = ||v_2||_1,
\end{equation}
that is, the initial drug load will be completely released. We will return to this point in section \ref{section:discussion}.

\section{Results}\label{section:results}
In this section we describe the implementation of our models outlined in sections \ref{section:model_1} and \ref{section:model_2}. All procedures have been implemented in \textsc{matlab} and programs will be available upon request from the corresponding author.

\subsection{The Lubachevsky-Stillinger protocol}\label{subsection:LS}

We work with a version of the Lubachevsky-Stillinger protocol \cites{Lubachevsky,Knott} that allows for classes of spheres with different radii. In the first step we convert the mass fractions of drug $f_D$, polymer $f_P$ and excipient $f_X$ into particle fractions, using the known particle radii $r_D,\,r_P$ and $r_X$ and the densities $\varrho_D,\,\varrho_P$ and $\varrho_X$ of the pure powders. For $i\in\{D,\,P,\,X\}$ let $m_i = r_i^3\varrho_i$ and set
\begin{equation*}
\tilde{n}_D =  1, \quad \tilde{n}_P =   \frac{f_Pm_D}{f_Dm_P} , \quad \tilde{n}_X = \frac{f_Xm_D}{f_Dm_X} ,  
\end{equation*}
and 
\begin{equation}\label{molar_fraction}
n_i =  \frac{\tilde{n}_i}{ \tilde{n}_D + \tilde{n}_P +\tilde{n}_X},  
\end{equation}
these are the particle fractions of drug, polymer and excipient in the powder mixture. The particle sizes, densities and mass fractions for different mixtures of the drug indomethacin (a commonly used anti-inflammatory drug), the polymer Eudragit RLPO and the excipient mannitol are given in table \ref{Tab1}. The corresponding particle fractions are given in table \ref{Tab2}.

A difficulty in the implementation of the Lubachevsky-Stillinger protocol is the decision when to stop. In theory, the time between two collisions (either sphere-sphere or sphere-wall collisions) approaches zero and hence the increase in the packing fraction approaches zero. We stop the execution if the times between successive collisions $t_{k+1}-t_k,\,t_{k+2}-t_{k+1},\,\dots\,,t_{k+n}-t_{k+n-1}<\epsilon$, for a number $n$ and a constant $\epsilon>0$ that are set by the user. The role of $n$ is to avoid premature termination if accidentally a time between two collisions is very short. In our implementation we choose $\epsilon=10^{-6}$ and $n=50$. This results in packing fractions of $\approx 0.54$ for packings of spheres with roughly equal radii (figure \ref{3D_pack}). This value is somewhat below $0.64$, the ``generally accepted'' packing fraction for a random dense monodisperse packing \cite{Torquato}. However, we consider it sufficient in this preliminary work. 

\begin{figure}[th]
\begin{center}
\includegraphics[width=60mm]{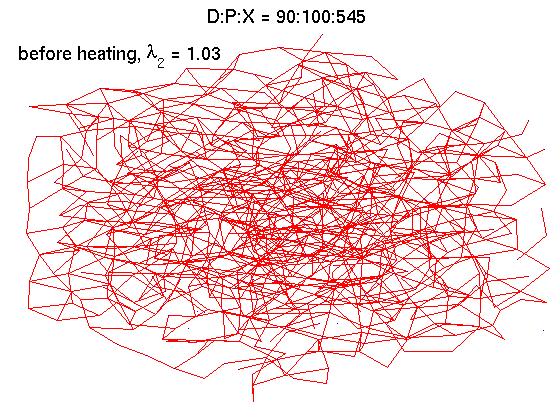}
\includegraphics[width=60mm]{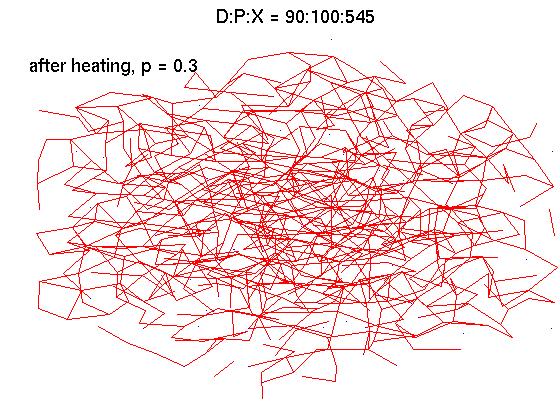}
\caption{The proximity graphs on the sphere centers of a random dense sphere packing. The constants in equations \eqref{connectedness} and \eqref{exterior} are chosen $\lambda=\mu=1.03$. The left panel shows the contact graph $G$ before the heating process. The edges are removed with a probability $p=0.3$ during the heating process, this results in the heated contact graph $\hat{G}$ on the right.}\label{adjacency_graph}
\end{center}
\end{figure}

At the termination of the Lubachevsky-Stillinger protocol at time $t^*$ we obtain a set of positions $(\boldsymbol{x}_i(t^*))_{i=1}^N$ and radii $(r_i(t^*))_{i=1}^N$. After possible compression of this sphere packing, we define the graph $G$ by joining vertices $\boldsymbol{x}_i$ and $\boldsymbol{x}_j$ if
\begin{equation}\label{connectedness}
|\boldsymbol{x}_i-\boldsymbol{x}_j| \le \lambda (r_i+r_j)
\end{equation}
where $\lambda \ge 1$ is a constant. Graphs of this kind of graph are also known as \textit{proximity graphs} on random point sets \cites{Meester,Penrose}. The heating process removes edges with a probability $p$ and results in the heated contact graph $\hat{G}$. Figure \ref{adjacency_graph} shows a contact graph of a packing before and after heating. Choosing $\lambda$ slightly greater than $1$ takes into account that the molecules can travel through interstitial spaces between particles that are close, but not touching. The spheres that have at least one surface point  close to a boundary of the domain are given the ``exterior'' label. For example, if
\begin{equation}\label{exterior}
\boldsymbol{x}_j^x \le \mu r_j  \quad \textrm{ or }  \quad \boldsymbol{x}_j^x \ge 1-\mu r_j
\end{equation}
for a constant $\mu\ge 1$, then that sphere is a possible point of exit in the $x$-direction. 

\begin{table}[th]
\begin{center}
\begin{tabular}{c|c|c|c|c}
$f_D$ & $f_P$ & $f_X$ & $h_1\, (mm)$ & $h_2\, (mm)$ \\
\hline
$10$  &  $10$ & $80$ & $3.06$ & $2.80$ \\
$10$  &  $20$ & $70$ & $3.09$ & $2.89$ \\
$10$  &  $30$ & $60$ & $3.16$ & $2.95$ \\
$10$  &  $40$ & $50$ & $3.19$ & $3.03$ \\
$10$  &  $50$ & $40$ & $3.24$ & $3.09$ \\
\end{tabular}
\vspace{\baselineskip}

\begin{tabular}{l|c|c|c}
 & drug & polymer & excipient \\
\hline
diameter ($\mu m$) & $100$  &  $100$ & $106 $ \\
density ($g/ cm^3$) & $1.37$  &  $1.23$ & $1.52$ \\
diffusion coefficient ($cm^2/s$) & $7\cdot10^{-6}$  &  - & $7\cdot10^{-6}$\\
\end{tabular}
\bigskip
\caption{Composition of tablets, particle sizes and true densities for drug (indomethacin), polymer, and excipient (mannitol or lactose). The fractions are the mass fractions in $\%$ in a tablet of total weight $500\,mg$. The heights $h_1$ and $h_2$ are those of a tablet of $13\,mm$ diameter at compression pressures of $74\,MPa\,(h_1)$ and $221\,MPa\,(h_2)$, respectively. The diffusion coefficients of drug and excipient are those in water. }\label{Tab1}
\end{center}
\end{table}

\begin{table}[th]
\begin{center}
\begin{tabular}{c|c|c }
$n_D$ & $n_P$ & $n_X$  \\
\hline
$12.2$  &  $13.6$ & $74.2$ \\
$11.7$  &  $26.1$ & $62.2$ \\
$11.3$  &  $37.6$ & $51.1$ \\
$10.8$  &  $48.2$ & $41.0$ \\
$10.4$  &  $58.0$ & $31.6$ 
\end{tabular}
\bigskip
\caption{The particle fractions of drug, polymer and excipient (in $\%$) corresponding to the mass fraction in table \ref{Tab1} are computed as in equation \eqref{molar_fraction}.}\label{Tab2}
\end{center}
\end{table}

\subsection{Release curves obtained from random walks on the contact graph}\label{subsection:random_walk}

On the heated contact graph $\hat{G}$ we perform random walks that start from each drug particle and end when an exterior vertex is reached, or the maximum number of steps, $N_{max}$, is achieved. As explained above, the probability of traversing edge $E_{ij}$ is  determined by the type of the terminal vertex,
\begin{equation}\label{propens}
c_{ij}=\left\{ 
\begin{array}{rl}
c_-  & \text{ if } j  \text{ is of type } P, \\ 
c_+  & \text{ otherwise}
\end{array}
\right. ,
\end{equation}
we use here $c_+= 100$ and $ c_-=1$. Notice that the current position is also a possible terminal vertex, that is, the random walker can remain in a fixed place for (perhaps long) periods of time. For every path we record the number of steps in the path and the length of the edges traversed. 
If a particle remains at a vertex, we still add ``one'' to the number of steps, and we add the diameter of the current vertex' particle type to the path length. Thus, even though a molecule may be close to where it started after one time step, the time step is still counted as ``time spent before exiting".  

\begin{figure}[th]
\begin{center}
\includegraphics[width=127mm]{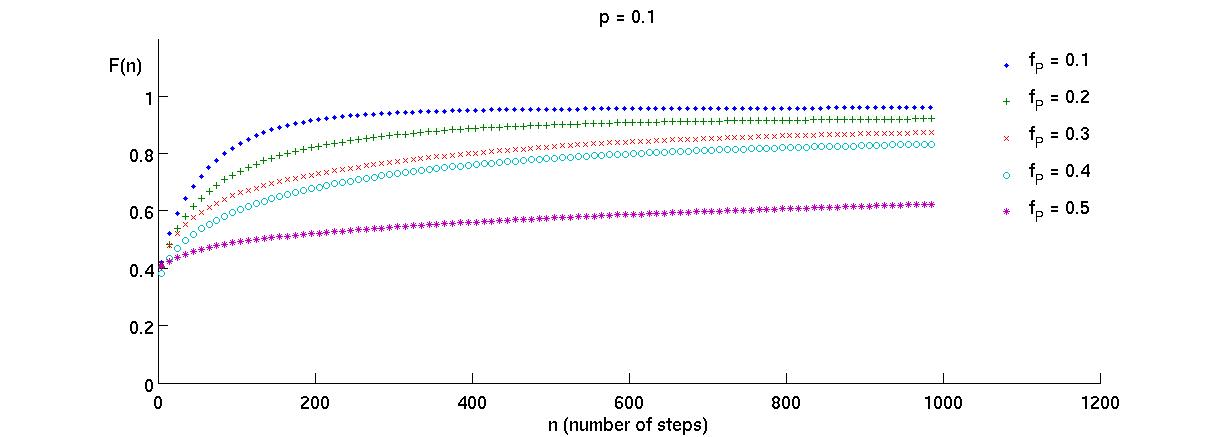}
\caption{Five release profiles obtained from step numbers of random walks as described in section \ref{subsection:random_walk}, as the polymer mass fraction varies. The mass fraction of drug is $10 \% $ throughout and their number in each packing is 180. For the actual particle fractions of drug, polymer and excipient see the left part of table \ref{Tab2}. We have used $\lambda=\mu=1.03$ in equations \eqref{connectedness} and \eqref{exterior} and a probability for edge removal $p=0.1$. The edge propensities are given in equation \eqref{propens}. The maximum number of steps is $N_{max}=10^3$.}\label{low_temp}
\end{center}
\end{figure}

\begin{figure}[th]
\begin{center}
\includegraphics[width=127mm]{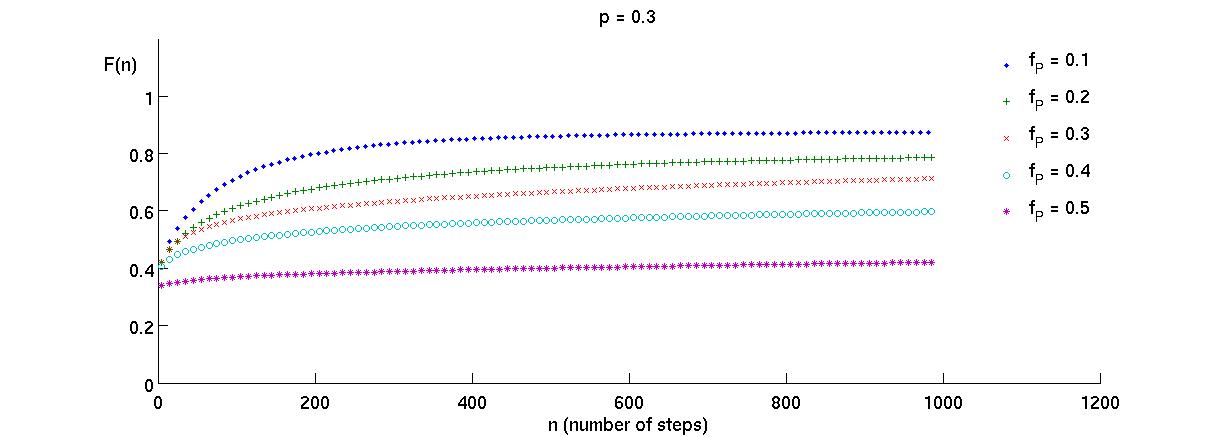}
\caption{As in figure \ref{low_temp}, but now with a probability for edge removal $p=0.3$.}\label{high_temp}
\end{center}
\end{figure}

We present simulated release profiles in Figures \ref{low_temp} and \ref{high_temp}. In these figures we vary the polymer fraction and keep all other parameters constant. We observe that as the polymer fraction increases, the drug release is slowed down. Further, we test different edge removal probabilities ($p=0.3$ and $p=0.5$) that can be interpreted as different curing temperatures. An increase in $p$ decreases the fraction of drug released.

\subsection{Release curves predicted by the continuous model}\label{subsection:pde_results}
\begin{figure}[th]
\begin{center}
\includegraphics[width=90mm,height=60mm]{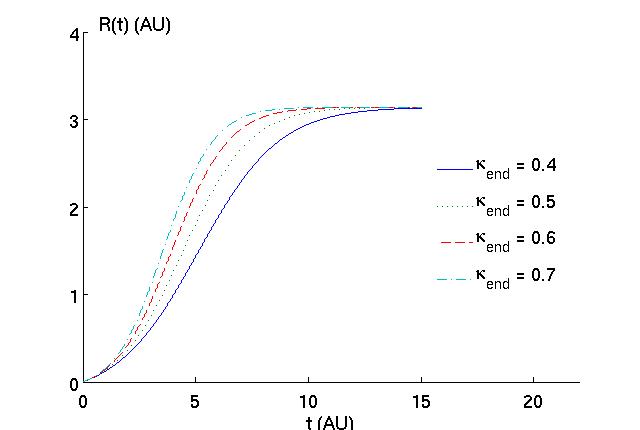}
\caption{Release profiles predicted by the continuous model \eqref{uv_system} as the final porosity $\kappa_{end}$ varies. The dimensionless parameters used in this example are $\kappa_0=0.02$, $D_u=0.3$, $D_v=0.5$ $\alpha_u=\alpha_v = 1.5$ and $C^u_{\max}=C^v_{\max}=u_2^0=v_2^0=1$.}\label{pde_release}
\end{center}
\end{figure}

As a first simplification we disregard the height of the tablet and collapse it to a disk, so that functions are now only dependent on the variable $r\in[0,1]$. We split the numerical solution procedure into a diffusion step for $u_1$ and $v_1$ and a reaction step for all four concentrations. It is assumed that during the diffusion step the porosity $\kappa(u_2+v_2)$ does not change. We discretize the radial Laplace operator on the uniform grid $r_k = (k-1) \Delta r,\, k = 1,\dots,N$ using standard finite differences and we use the Crank-Nicolson scheme at every diffusion step \cite{Press}*{Section 17.3}. The reaction step is calculated using the Euler forward method. Release curves obtained from numerical solution of equation \eqref{uv_system} are shown in figure \ref{pde_release}. A larger final porosity results in a faster release of the drug. Observe that all release curves in figure \ref{pde_release} predict a complete release of the drug, leading to the conjecture \eqref{pde_complete_release}.

\subsection{Comparison with experimental release curves}\label{subsection:exp_data}
\begin{figure}[th]
\begin{center}
\includegraphics[width=100mm,height=60mm]{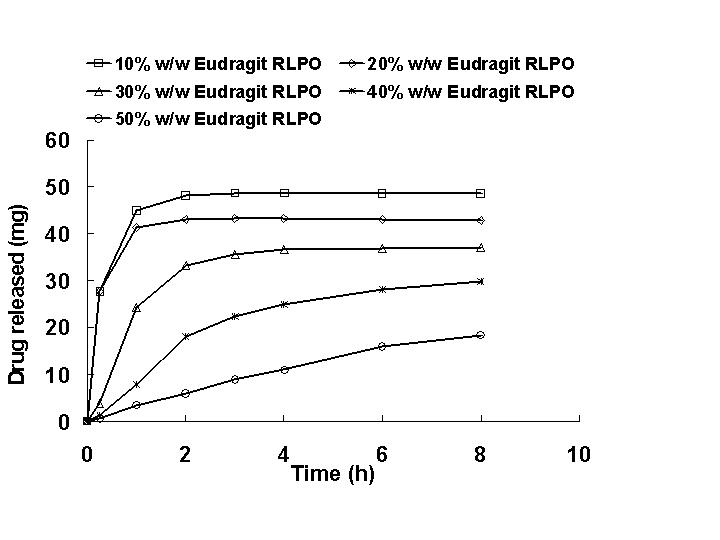}
\vspace{-10mm}
\includegraphics[width=100mm,height=60mm]{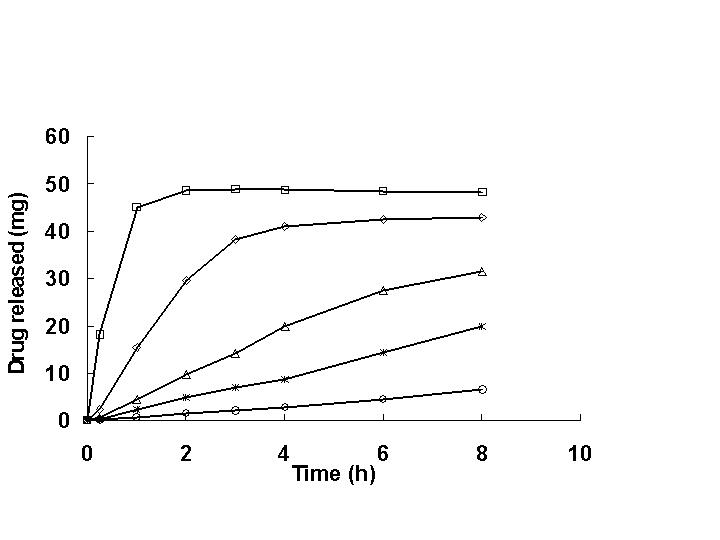}
\caption{Release of indomethacin (mass fraction $10\%$) from Eudragit RLPO matrix tablets containing mannitol ($90-125\,\mu m$ particle diameter), a plastic excipient, using the USP basket apparatus at $100\, rpm$ in $900\, ml$ phosphate buffer $pH= 7.2\, (0.2\, M)$ at $37{}^\circ\,C$. The top figure shows a tablet that was compressed at $74\,MPa$ and cured $24\,h$ at $40{}^\circ\,C$, the bottom figure shows a tablet that was cured  $24\,h$ at $70{}^\circ\,C$. Data are representative for three repetitions of the experiment (from \cite{Chatterjee}). }\label{exp_data}
\end{center}
\end{figure}

Experimental tablets were formulated from mixtures of indomethacin, Eudragit RLPO and mannitol and their release profiles were determined, see \cite{Chatterjee} for a detailed description. Briefly, powder mixtures were compressed at different pressures and tablets were heated at selected temperatures. The tablets were then placed in a phosphate buffer medium and samples were collected at different time points over a period of $8\, h$ (see figure \ref{exp_data}). We observe a good qualitative agreement with our simulated release profiles in figures \ref{low_temp} and \ref{high_temp}.

\section{Discussion and Conclusion}\label{section:discussion}

Several studies have investigated percolation on regular lattices and graphs obtained from random sphere packings. Villalobos and collaborators \cites{Villalobos05,Villalobos06} used Monte Carlo simulations on cubic lattices to predict drug release profiles from binary drug/excipient mixtures. In these works the excipient played the role of the inert matrix, which in our case is the polymer. It was reported that if the particle fraction of the inert matrix (the inaccessible sites in the lattice) is below a threshold of $\approx 69\%$, then the release of the drug is slowed down with increasing matrix fraction, but will still be complete. Only above the critical matrix concentration an entrapment of drug will be observed. The value of $69\%$ is complementary to the estimated site percolation threshold of $0.3116$ for the cubic lattice \cite{Sur}. A similar value was reported by Powell \cites{Powell1979,Powell1980} for percolation on the contact graph of a monodisperse random sphere packing. In this work we include the heating of the tablet, which melts the polymer. The blocked polymer particles and the removal of certain edges provide a case of what is known as site-bond percolation \cite{Chang}. In their paper \cite{Chang}, Chang and Odagaki studied site and bond percolation processes where sites and bonds are removed independently from each other and eventually open sites are connected if they share an open bond. They found that for the cubic lattice, the percolation threshold as a function of the probabilities that sites respectively bonds are open, is well described by a hyperbola. While this was not the primary goal of our study, we ended up addressing the problem of determining percolation thresholds for contact graphs of dense random sphere packings. 

While we have seen a good qualitative agreement between our simulated release profiles and experimental data, further research is needed to provide more reliable and quantitative predictions.  In our simulations we find many particles with very short escape paths. However, their number is going to diminish as we consider sphere packings with more spheres. Intuitively, the fraction of spheres ``close to the exterior'' is decreasing as the number of spheres in the packing increases, a problem that was discussed already in  \cite{Powell1979}. Larger simulations, perhaps using parallel computing methods, are required.  Further work will also include careful calibration of the parameters $\lambda, \,\mu$ and $p$ to experimental release curves. 

Our continuous model captures nicely the change of the release curves from convex to concave. However, as figure \ref{pde_release} and equation \eqref{pde_complete_release} suggest, the release will always be complete. No percolation behavior is exhibited by the system \eqref{uv_system}. In the future we need to take into account the permeability of the porous matrix, a notion that arises in hydrology \cite{Bear}. In the derivation of the dissolution kinetics \eqref{dissolution_kinetics}, we have assumed that the dissolution of drug and excipient is helped by an increase in porosity. This is contrary to the commonly made assumption of a receding boundary in the pharmaceutical literature \cite{Higuchi}. There, the drug dissolves in a way that decreases the area of contact between drug and solvent. It poses an interesting inverse problem to distinguish these two concepts, given the experimentally determined release profiles. 

Another interesting suggestion of Villalobos \textit{et al.} is that the release profiles have a
Weibull cumulative distribution function,  \cite{Villalobos06}*{Equation 2}.
\begin{equation*}
 \frac{F(t)}{F_\infty} = 1-\exp(-at^b)
\end{equation*}
where the parameters are influenced by the size of the matrix tablet. More precisely, the ratio of the number of exit sites versus the total number of sites, $N_{exit}/N$ should influence the release kinetics, with faster release for smaller values of  $N_{exit}/N$. If we assume that $N_{exit}\propto N^\frac{2}{3}$ then larger tablets result in slower release. 
While this makes sense intuitively, we consider this to be an interesting question to explore more rigorously, and  we plan to address it in future work.

\section*{Acknowledgments}
The authors would like to thank the School of Pharmacy at the University of Otago for its hospitality and in particular the organizer, Dr.~Sarah Hook for making the workshop in  December 2007 possible. We thank Mousab Arafat, Chris du Bois, Emma Spiro and Bram Evans for participation in our working group and Tracy Backes for helpful suggestions on the implementation of the Lubachevsky-Stillinger protocol. 

\bibliography{problem4}

\end{document}